\documentclass{article}%
\usepackage{amsfonts}
\usepackage{amsmath}
\usepackage{amssymb}
\usepackage{graphicx}%
\providecommand{\U}[1]{\protect\rule{.1in}{.1in}}

\begin{document}

\title{A note on Infraparticles and Unparticles}
\author{Bert Schroer\\CBPF, Rua Dr. Xavier Sigaud 150, 22290-180 Rio de Janeiro, Brazil \\and Institut fuer Theoretische Physik der FU Berlin, Germany}
\date{October 2008}
\maketitle

\begin{abstract}
The observed astrophysical phenomenon of dark matter has generated new
interest in the problem of whether the principles underlying QFT are
consistent with invisibility/inertness of energy-momentum carrying "stuff" as
e.g. "unparticles". We show that the 2-dim. model which has been used to
illustrate the meaning of unparticles belong without exception to the class of
former infraparticles. In d=1+3 infraparticle are identical to electrically
charged particles which despite their nonlocality are our best particle
physics "candles". The "invisibility" in this case refers to the infinite
infrared photon cloud with energies below the resolution of the measuring
apparatus which can be made arbitrarily small by increasing the photon
registering sensitivity but not eliminated. This is not quite the kind of
invisibility which the unparticle community attributes to their invisible
"stuff" and whose existence would probably contradict the asymptotic
completeness property. The main aim of the present work is to show that
knowledge about this part of QFT is still in its infancy and express the hope
that the work on unparticles may rekindle a new interest in conceptually
subtle old unsolved important problems instead of inventing new once which
after some time increase the list of unsolved old ones.

\end{abstract}

\section{Previous incursions beyond the standard particle setting}

The quest for understanding the particle content of QFT beyond the standard
mass gap hypothesis (one-particle states separated by a finite distance from
the contiuum) has been an important topic for a long time. The study of these
problems began after it became clear that interactions, which become
sufficiently strong in the infrared regime, can and will change the conceptual
basis of standard scattering theory. The oldest model to address this issue is
the famous Bloch-Nordsiek model which, via the Yennie-Frautschi-Suura infrared
treatment of scattering theory of charged particles \cite{YFS}, led in the
early 60s to the first ideas about \textit{infraparticles} \cite{S1} i.e.
charged particles permanently surrounded by an infinite cloud of soft photons
below the visibility limit. The most recent attempt in this direction is
Georgi%
\'{}%
s proposal of "unparticles" \cite{Geo} which are thought to lead to an
"invisibility" of a certain kind of outgoing matter component, which shows its
presence through the appearance of energy-momentum-carrying but otherwise
undetectable "fractional particle stuff".

Unlike the infraparticle concept this so formulated
"unparticle"-problem\footnote{The terminology is taken from the papers and
probably refers to their presumed invisibility. The name "stuff" is also used
on several occasions in those papers.} is not imposed by any observational
fact; it is at this stage a mere mind game, although dark matter is sometimes
mentioned as a potential observational application. Clearly what is called
"stuff" by those authors is outside the standard particle world and its
conjectured appearance together with scattering of ordinary particles is part
of the larger "asymptotic completeness" problem i.e. of the question whether
it is possible that the particles which emerge asymptotically in a scattering
process do not form a complete set of states but leave some stuff which
dissipated into the vacuum in such a way that it cannot be accounted for in
particle registering devices. So any set of physical assumption which leads to
asymptotic completeness has a bearing on this problem; we will return to this
point at the end of the paper.

It is our intention to test the consistency of this idea within the setting of
local quantum fields and in particular to compare this unparticle proposal
with the problem to detect infinite\footnote{The presence of photon "clouds"
i.e. an infinite number of soft photons requires their exact masslessness.}
soft photon clouds around infraparticles\textit{. }The study of infraparticles
is a well researched subject which started in the 60s with the investigation
of certain two-dimensional models \cite{S1} and reached a certain amount of
conceptual maturity in the 80s, when it was shown that 4-dimenional
(electrically) charged particles are infraparticles \cite{Fro}\cite{Bu1}%
\cite{Bu2}.

Some of these old results will be reviewed; this is warranted because the new
development has apparently been taking place without much awareness about the
old achievements, This is of particular relevance since the task to explore
the field-particle relation beyond the boundaries of standard particle physics
(existence of discrete masses separated by gaps from the continuum) was
already the aim of the infraparticle investigations. By comparing the
presently still only half-baked unparticle idea with the more mature
infraparticle physics one hops to learn if and in what sense the former
represents a conceptually viable new trans-particle idea.

The unravelling of the relation between quantum fields and particles has been
one of the most difficult tasks ever since field quantizations was discovered
in the late 20s. The subtlety of this problem became first highlighted in the
work of Furry and Oppenheimer \cite{F-O} when these authors found that in
interacting QFTs every field, including the "fundamental" Lagrangian fields,
never creates a pure one-particle state (and not even a one-particle state
with a limited finite number of particle/antiparticle pairs), but its local
creation is always in company with an infinite particle/antiparticle
polarization "cloud"\footnote{In finite order of perturbation theory there is
a finite admixture which increases with the order and becomes a "cloud" in the
limit.} whose extension and shape depends on the kind of local interaction.
This observation followed in the heels of Heisenberg%
\'{}%
s discovery of a more mild form of vacuum polarization associated the
composites (Wick-products) of free fields while trying to define the "partial"
charge localized in a compact spatial region which is associated with the
spatial integral over bilinear current of a free complex scalar quantum field.

In a modern setting the F-O observation amounts to the nontriviality of the
(connected) formfactors of any local operator $A$ in an interacting QFT, which
by the crossing property\footnote{In standard QFT (pointlike interpolating
fields, mass shells with spectral gaps) the vacuum polarisation formfactor is
related by crossing to the formfactors with different distributions of in and
out particles in ket and bra states.} are all related to a an analytic master
function, usually identified with the vacuum polarization formfactor of $A$
\begin{equation}
^{out}\left\langle p_{1},...p_{n}\left\vert A\right\vert 0\right\rangle \neq0
\end{equation}
These formfactors, as a result of their crossing analyticity, fulfill a kind
of "Murphy%
\'{}%
s law", stating that all channels, whose coupling is not forbidden by charge
superselection rules, and their associated symmetries, are indeed coupled.
This is very different from QM (even in its relativistic form, as the theory
of direct particle interactions (DPI) \cite{interface}) where one can
couple/decouple channels at will by manipulating the interaction potentials.
Whereas this "no decoupling of channels" situation in QFT does not have the
status of a theorem which can be found in the literature but is certainly
consistent with the the nonexistence of any counterexample, there is the
weaker statement, the \AA ks theorem \cite{Aks}, showing that in $d\geq1+2$ a
QFT must have on-shell particle creation if it has has any nontrivial elastic
scattering at all. A breakdown of this kind of "benevolent Murphy's law" in
the mass gap setting is only possible in a setting which avoids the crossing
property. This in turn can only happens in theories in which at least some
generating massive fields are semiinfinite stringlike localized
\footnote{Buchholz and Fredenhagen proved that semiinfinite
string-localization is the most general possibility allowed by the mass gap
hypothesis in conjunction with the assumption of the existence of a point-like
generated neutral observable subalgebra \cite{Bu-Fr}.}. The problem with this
putative partial "on-shell blackening" is that no interacting demonstration
model has yet been found.

The problems related to these first (Furry-Oppenheimer) observations were
finally solved in the late 50s with a reasonably good first understanding of
the field-particle relation. Contrary to the Fock space formulation of QM,
interacting QFT connects fields with particles only through the asymptotic
large time limits of scattering theory. The derivation of incoming/outgoing
free fields (which lead to a Wigner-Fock space particle structure of standard
QFT) from the short range nature of the connected part of correlation
functions (a consequence of the mass-gap energy momentum spectrum and causal
locality) has been one of the high points in the conceptual understanding of
QFT, with profound experimental consequences.

The first necessity to go beyond this standard setting arose from the
observation that electrically charged particles do not fit into this framework
since the S-matrix, as represented by the on-shell restriction of time ordered
correlations, has infrared divergencies which cannot be removed by
renormalizing the parameters of QED. This divergencies are not a mere
consequence of the violation of the gap hypothesis. As the Yukawa coupling
between nucleons and massless pions which is infrared finite and fits
perfectly into the standard particle/field framework shows, the infrared
divergencies, which lead to a breakdown of the standard particle setting for
electrically charged particles, are the result of an increase of interaction
strength in the infrared of the coupling of photons to charged
fields\footnote{If the authors of the statement "Observed, known particle
physics is based on theories which have a mass gap and/or are free in the
infrared" \cite{Grin} really mean what they write, they would have missed out
on QED.}, it does not happen in the mention Yukawa coupling since no matter
how large the coupling paramter is, scalar or pseudoscalar couplings cannot
reach the necessary infrared strength. The Bloch-Nordsiek method and its
refinement in the work of Frautschi-Yennie-Suura \cite{YFS} shows that the
infrared stable quantities, which replace the standard scattering amplitudes,
are the inclusive cross sections in which the photons below a certain
resolution (which varies with the sensitivity of the measuring hardware)
escape undetected.

This led to a profound revision of the particle-field relation. The simplest
way in which this new particle aspect revealed itself was through a change in
the two-point functions: instead of the mass-shell delta function in the
Kallen-Lehmann two-point function of a physical (gauge invariant) electrically
charged field does has a "infraparticle" singularity which starts at
$p^{2}=m^{2}$ in a inverse power-like fashion and extends into the
multiparticle continuum. Unitarity (Hilbert space positivity) limits this
interaction-dependent power to be milder than the mass shell singulariy. This
in turn leads to a vanishing\footnote{It is not uncommon that an object which
diverges in perturbation theory vanishes on structural grounds.} large time
asymptotic LSZ limits, thus underlining the breakdown of standard scattering theory.

There has been steady progress in a nonperturbative structural understanding;
the strongest result, a milestone in the conceptual conquest of infrared
aspects of QFT, has been obtained by Buchholz. He showed that an appropriate
formulation of the quantum Gauss law \cite{Bu2} is incompatible with the
standard particle structure. The infraparticle structure, including the
spontaneous breakdown of Lorentz invariance in electrically charged states, is
a consequence of this observation.

The derivation of inclusive scattering formulas, which bypasses amplitudes and
produces directly inclusive probabilities, remained however a still
incompletely achieved goal. Although one believes to have all the relevant
concepts in place \cite{Por}, these attempts did not really lead to useful
nonperturbative formulas for inclusive scattering probabilities which can
match the formal elegance of the standard LSZ formulas.

A related problem, raised by the unparticle community is the question of the
possible particle manifestation of conformal QFT. Here the LSZ limits of any
field with noncanonical (anomalous) dimension again vanish, as in the case of
infraparticles. But in addition one finds a stronger theorem stating that any
conformal field with a canonical (free field) short distance behavior is
necessarily itself a free field \cite{S2}. Every field has a scaling limit,
but not all such limiting theories are conformal. On the other hand each
conformal theory is believed to arise through a scaling limit of a standard
theory. This raises questions about the possible particle physics aspects
which can be extracted from a conformal theory.

Although according to the best of my knowledge there are no theorems, one
believes that only highly inclusive cross sections can counteract the
vanishing of the LSZ limits. Some people have pointed out that this may not be
enough; in order to arrive at finite cross sections one should generalize the
inclusiveness aspect as an averaging procedure over incoming
configurations\footnote{Such doubly inclusive cross setions play a role in the
Kinoshita-Lee-Nauenberg theorem \ \cite{K-L-N}.}, With other words, one
expects such possibly "doubly inclusive cross sections" in a standard massive
theory to remain finite in a scaling limit in which the inclusivness
(resolution energy) parameters remain fixed. There seem to exist no proofs for
these claims of relating inclusive cross sections with conformal QFT. Whereas
the answer concerning the simple inclusive cross sections seems to be negative
(infrared divergent), the feasibility of the double K-L-N appears still open.

Even though the informed reader will recognize an identity of motivation since
both the infraparticle setting and the unparticle idea aim at conquering the
ground beyond the standard particle theory, there is one difference with
respect to the issue of "invisibility". The infrared "stuff" which constantly
oozes out from an infraparticle (the infinitely many soft photons below an
energy resolution hover with nearly infinite extension around an electrically
charge particle) is not invisible per se. The hard photons with energies above
the resolution coming out of a scattering process involving charged
infraparticles are very visible and the "invisibility" in the infrared is
determined by outside resolution parameters on the registration side, although
there is always an infinite cloud remaining which escapes detection. On the
other hand, since the protagonists of unparticles do not seem to think in
terms of resolving the hard component of their stuff, the only remaining
possibility is that their unparticle stuff is intrinsically invisible,
independent of the infrared strength of the interaction. But if this is the
case, what role remains for the infrared properties of interactions as the
cause of "invisibility".

The next section contains some details on infraparticles which up to the
present constitute the only known mechanism to transcend the limitation of
standard particle physics. At the end of that section the reader is expected
to understand why a critique of unparticles requires a good understanding of infraparticles.

\section{A brief anthology of infraparticles}

A conceptual understanding of infraparticles and their scattering in the
realistic 3+1 dimensional case of charged particles in QED is much more
difficult than the standard scattering theory of massive particles in the
presence of a mass gap. Already in second order perturbation theory infrared
divergencies arise in the scattering amplitudes. These divergencies cannot be
absorbed into a renormalization of physical parameters. Although the
suggestion that this indicates a \textit{conceptual change in the notion of
particles} is quite old, its concretization in the setting of QED turned out
to be a long lasting scientific endeavor.

The infraparticle idea was first exemplified and tested in the "theoretical
laboratory" of 2-dimensional models \cite{S1}. Those models which were known
in the 60s\footnote{Besides the mentioned model \cite{S1} list of those models
consisted of the (massless) Thirring model, the Schwinger model and variations
and combinations thereof.} had no genuine interactions in the sense of
scattering, but they certainly contained interesting messages about a modified
particle structure which could account for the infrared divergencies observed
in collisions in which electrically charged particles participate \cite{YFS}.

These models strongly suggested that it was the modification of the particle
structure whose nonobservance led to the divergencies. But the change from
particles to infraparticles also required to abandon the standard setting of
scattering amplitudes and pass to inclusive scattering \textit{probabilities}%
\footnote{"Inclusive" meant that infrared photons below a certain sensitivity
$\Delta$ of the registering apparatus were summed over.}. It was found that
the logarithmic infrared divergences in on-shell perturbative amplitudes of
charged particles, calculated according to the standard rules, are compensated
with divergencies in multi photon creation contributions to inclusive cross
sections; the remaining finite terms summed up to an interaction dependent
power law for small values in the inclusive resolution parameter.

This model observation suggested that the radical change of the particle
concept amounts in momentum space to an amalgamation of the particle
mass-shell with the continuum which, unlike the ubiquitous vacuum polarization
contributions in any interacting QFT, cannot be gotten rid of by the large
time asymptotics of scattering theory. But if the standard scattering theory
breaks down, then the Hilbert space of the model does not have the form of a
multiparticle Fock space. Indeed the "exponential massless Boson fields" which
appear at first in \cite{S1} (and later in all the other 2-dim. infraparticle
models), do carry a superselected charge and therefore cannot live in the
bosonic Fock space defined by the creation/annihilation operators.of the
well;defined current (derivative of the Boson field). Rather they generate a
bigger algebra living in a bigger Hilbert space of which the chargeless
algebra generated by the currents is a subalgebra and the original Fock space
is a subspace of a space of charged states without a Fock space structure.

This change should reveal itself in the K\"{a}ll\'{e}n-Lehmann representation
as a modification of the mass-shell contribution and indeed this was precisely
what one observed in the very first calculation \cite{S1} where instead of the
mass-shell delta function one found a milder singularity in form of a cut
starting at $p^{2}\geq m^{2},~p^{0}\geq m.$ In the setting of operators the
basic field of this model (the derivative coupling of a massless scalar to a
massive spinor) was described by a product of a free \ massive Dirac field
with an exponential of a zero mass scalar Bose field i.e.%
\begin{equation}
\psi(x)=\psi_{0}(x):e^{ig\varphi(x)}: \label{field}%
\end{equation}
There is a subtle point concerning the precise meaning of this exponential
since as mentioned before the massless scalar boson $\varphi$ itself is not to
be considered as a bona fide pointlike object\footnote{This "field operator"
is really an operator-valued distribution whose testfunctions space is
restricted to those Schwartz test functions whose total integral vanishes. Its
Hilbert space is generated by polynomials in the well-defined associated
current (its derivative, the field is a line integral in terms of this
current).} and the exponential is amalgamated with the free Dirac field in
such a way that the Hilbert space of the model does not contain a $\psi
_{0}(x)$ Fock subspace.

There are different ways to make this point explicit. The simplest is to start
from an exponential of a massive two-dimensional free field, which lives in
the Fockspace of the free field\footnote{This property is lost in the zero
mass limit when the free field diverges in the infrared (but its derivatives
stay finite) and the Wick rules for exponentials suffer restrictions from the
charge conservation.} and hence obeys the unrestricted Wick contraction rules,
and to perform a zero mass limit within the vacuum expectation values. In the
massless limit the exponential operators in the correlation function must be
multiplied with a certain anomalous power in the mass which is chosen such
that no correlation diverges, but not all vanish. The power needed is this is
the same as given by a formal scaling argument.

Another method is to work in the Hilbert space of currents and define the
desired exponentials as a limit of a bilocal line integral of the current with
one end going to infinity; doing this inside the correlation functions leads
again to the previous result.

The massive free field in 2 spacetime dimensions has no other physical
representations than the usual charge-less vacuum representation. Its massless
limit is however very special in that it leads to the only free field with
continuously many charged representation formally generated by the exponential
function, a fact which was noticed already by Jordan\footnote{Unfortunately
Jordan used his correct observation of what we nowadays would call
Bosonization/Fermionization to base his pet idea of "the neutrio theory of
light" on \cite{Jor}. The relation of the gauge invariant content of the
Schwinger model in terms of a exponential massive free Boson which has no
charge sectors to its short distance limit with infinitely many charge sectors
is an impressive allegory for the transition from confinement to short
distance charge liberation.}. However his attempt to sell this observation
under the heading \ "neutrino theory of light" (believing erronously that this
can be generalized to higher dimensions) was not very successful; it led to a
very funny mocking song composed by his colleagues \cite{Pais}.

A field which is a local function of free fields (and hence is local relative
to free fields) has no interaction (no scattering), even though its
correlation function (even its two-point function) look like anything but
free. This also holds true for fields which result from a charge generating
scaling limit procedure from free fields, as the above exponential.

What can and does happen through the use of such charge-carrying exponentials
however, is that the Hilbert space obtained from the reconstruction using the
limiting correlation functions is different from the Wigner-Fock space of the
original particles. It is easy to see that the presence of the charge-carrying
exponential modifies the mass-shell delta function into a fractional power
(not a fractional number of particles as the unpartcle partisans claim) in
terms of the K\"{a}ll\'{e}n-Lehmann spectral variable $\kappa^{2}$, in short
its defines an "infraparticle"\footnote{The only restriction on its range
comes from the requirement that the exponential acts in a Hilbert space which
follows from the mentioned charge superselection rules.}. The field
(\ref{field}) originates from a Lagrangian which describes a conventional
derivative coupling of a two-dimensional massless scalar with a massive
spinor. All other soluble models (including those which have recently been
used to explain the notion of unparticles \cite{Geo2}) see later) of the 60s
and 70s, with the exception of the unmodified Schwinger model\footnote{It was
shown in \cite{L-S} the gauge invariant content of the Schwinger model is an
exponential of a massive scalar field. As mentioned before, this exponential
passes to a charge-carrying exponential zero mass operator. This illustrates
the (gauge-invariant) "Schwinger-Higgs charge sceening" with the unscreened
charges appearing in the "asymptotic freedom limit". It also illustrates a
peculiarity of 2-dim. free fields which already played a role in Jordan's
"neutrino theory of light" \cite{Jor}.}, have these charge-carrying zero mass
exponential factors. The local observables of these models always contain the
current operator $\partial\varphi,$ whereas the exponentials are charge
creators in the mentioned sense i.e. objects which interwine the different
superselection sectors of the respective models.

The testfunction-smeared infraparticle operator applied the the vacuum yields
a state which captures much more of the localized testfunction than just the
mass shell restriction of its Fourier transform which the free field was able
to extract. As a result the separation into a "particle" like contribution and
the remaining "stuff" is not as well-defined as in standard particle theories.
Note also that if one attributes to the word "stuff" the meaning of an
uncountable substrate, it is not the emitted higher frequency photons (which
enter the registring device), but rather the invisible uncountable long range
part which deserved the predicate "stuff" and is at least partially (below the
resolution on the observer side which can never be completely eliminated) "invisible".

It is not uncommon that what is infrared-divergent in perturbations theory may
sum up to be zero nonperturbatively. Indeed the LSZ limits of infraparticle
fields as (\ref{field}) are zero since Hilbert space positivity forces the
mass-shell singularities to be milder than a delta function. This means that
standard scattering theory is not applicable to infraparticles, but the
objects beyond the standard setting are not necessarily "invisible". A
calculationally efficient formalism to compute inclusive cross section for
infraparticles exists only in a rudimentary fashion \cite{Por}; the most
efficient method is still the Yennie-Frautschi-Suura infrared
regularization-based compensation method which in turn is a generalization of
the Bloch-Nordsieck formalism.

The conjecture, based on the change of the mass-shell structure of the
Kallen-Lehmann two-point function in those models, that the cause for the
breakdown of the standard scattering theory was a rather radical change of
notion of particles, was a bit audacious in the 60s. Merely viewing this
change in the analytic setting of poles and singular cuts as a singular cut
replacing the delta function in the K-L spectral function, would not reveal
the full dynamical structure of infraparticles. What was needed was an
understanding in terms of \textit{spacetime localization properties}.

Partial results about the realistic case were found in \cite{Fro}, and a more
complete conceptual picture emerged in \cite{Bu1}\cite{Bu2} (see also
\cite{Por}\cite{Haag}) where a theorem was proven according to which the
infraparticle structure together with the spontaneous breakdown of the
L-symmetry in charged sectors (related to the infinite cardinality of the
infrared photon clouds) is a consequence of the appropriately formulated
quantum version of the Gauss law. This limits the infraparticle nature to
abelian gauge theories, but represents nevertheless (in my opinion) a high
point for what can be achieved by rigorous structural arguments.

There remains a practical question namely how does a physical charge-carrying
operator look like? From Buchholz's theorem we can conclude that it must be
extended up to infinity. The formal candidate which has the sharpest
localization which is consistent with the Gauss law is of the
Dirac-Jordan-Mandelstam form%
\begin{align}
\Psi(x,e)  &  =~"\psi(x)e^{\int_{0}^{\infty}ie_{el}A^{\mu}(x+\lambda
e)d\lambda}"\label{DJM}\\
\Psi(x,e)  &  \rightarrow D(\Lambda^{-1})\Psi(\Lambda x,\Lambda e)\nonumber
\end{align}
where $e$ is a spacelike unit vector which characterizes the localization
along the line $x+\mathbb{R}_{+}e$ and the electric charge is denoted by
$e_{el}$. This expression fulfills all the formal requirements. It is gauge
invariant and extends to infinity in accordance with Gauss law. It is a
string-localized field which transforms covariantly (second line) but the
L-invariance is spontaneously broken i.e. the implementing global unitaries of
the algebraic automorphism do not exist \cite{Bu2}. It is an interesting and
poorely understood question whether such formulas are a necessary structural
consequence of the nature of the local observables. Combining an old idea of
reconstructing charged fields from neutral currents by using a lightlike limit
procedure which Langerholc and myself designed in the 60s \cite{La-Sc} Jacobs
\cite{Jacobs} introduced the concept of gauge bridges and showed that at least
in the abelian case and in the quasiclassical approximation of QED the above
formula is canonically distinguished in the sense that it can be obtained in a
natural way from local observables only. Unfortunately it is not clear whether
this holds also for the quasiclassical approximation of the QCD model.

Another unexpected but related feature was that the different spacelike
directions, which after smearing with directional functions $g(e)$ with small
support become narrow spacelike cones, are defining superselection rules in
addition to those of the electric charge (or the electric charge is the
directional-independent part of a finer superselection structure). The
physical mechanism behind is that these cones contain an infinite accumulation
of soft photons which makes it impossible to pass from one cone direction to
another by a local or at least quasilocal change \cite{Haag}.

It takes tremendous computational stamina to proof that this formal expression
(\ref{DJM}) admits a renormalized version in every order of perturbation
theory, but exactly this was accomplished by Steinmann\footnote{The
computational effort necessary to assure the perturbative existence of these
DJM formulas goes beyond what any standard renormalization formalism as
\cite{Salam}\cite{Weinberg} or any of the more recent refinements can achieve.
Steinmann had to develop a technique especially for this problem.} \cite{St}.
There are of course other noncovariant ways of organizing the localization in
accordance with Gauss law as e.g. a Coulomb-like distribution which is
rotationally invariant around x in a fixed reference frame, but the
semiinfinite string localization which represents a singular limit of a
spacelike cone is the best analogy to the point as the sharpest limit of a
causally closed (i.e. double-cone shaped) compact region and in the sense of
maintaining Poincar\'{e} covariance. Note that the line integral in the
exponential corresponds to the zero mass scalar field in the 2-dim. setting in
that its perturbative modifications in the exponential interaction strength
also lead to momentum space logarithmic corrections (a power law modification
after summation). This is in accord with the two-dim. exponential massless
Boson calculation in \cite{S1} and \cite{Geo2}, with the only difference that
in the above case the gauge invariance prevents to attribute a separate
Hilbert space meaning to the two factors in (\ref{DJM}).

The spacetime analysis of infraparticle is not only more intricate than the
study of the infraparticle structure of the two-point function near
$p^{2}=m^{2},$ it is also much more revealing. For example it would be
virtually impossible to conclude from the changed mass shell singularity
structure of the two point function alone that the sharpest localization of
infraparticles is semiinfinite spacelike and that Gauss's Law is the cause of
all these modifications.

Behind the esthetic flaw of having to do things "by hand" instead of getting
them from the perturbative formalism as all the other expectations of
pointlike fields, there exists a problem which becomes much more pressing in
QCD, where \textit{no consistent formula "by hand" for nonlocal gauge
invariant operators which corresponds to (\ref{DJM}) has been found}. This is
of course related to the problem of gluon- and quark- confinement and possibly
of dark matter (in the sense of matter which is largely inert with respect to
standard matter but nevertheless appears to coexist in the same theory).

Such an "invisible" counterpart of the charged QED matter, if it exists,
cannot be understood as part of the existing gauge theoretical formalism
aiming at local observables which are identified as the gauge invariant part
within an unphysical setting. Possibly nonlocal gauge invariants in QCD are a
fortiori not part of the formalism but left to ingenious guesswork. Whereas in
the above abelian case of physical charged fields this was still possible on a
formal level (\ref{DJM}) as well as under the more stringent conditions of
renormalized perturbation , nothing is known about nonlocal operators in a
physical Hilbert space in QCD-like models except those vague ideas associated
for the last 4 decades with confinement of quarks and gluons\footnote{The
putative link between asymptotic freedom and infrared slavery has the flaw
that it does not account for all degrees of freedom which were initially
there.} which draw their main support come from placing quantum mechanical
quarks into a vault created by the walls of a potential or from lattice gauge
theory which is not even able to predict the simpler infraparticle properties
of QCD. QFT does not dispose over such resources, contrary to the quantum
mechanical vault mechanism its very restrictive causal locality principle only
leave the infinite spacetime extension as the resource of "invisibility" of
certain matter components. This resource was already used in a weak form by
the undetected infrared photon component, but as mentioned, it is not a
consequence of the presence of zero mass particles alone, one also needs an
interaction which is sufficiently strong in the infrared; the $N$-$\pi$
interaction with massless $\pi$ does not have the strength to create infrared clouds.).

The problem starts when a zero mass gluon acts on itself. In a metaphoric
picture interacting should inherit the partial invisibility of infrared
photons, but on the other hand they are also required to behave like charged
infraparticles whose "least nonlocal" localization is a semiinfinite string as
(\ref{DJM}) i.e. they have to be the source and that what it produces at the
same time. How can these two tendencies be reconciled in a non-metaphoric way?
It seems to me that the first step in this direction should be to look for a
reformulation which loosens the shackles of gauge theory to local observables
and get local and nonlocal observables (="nonlocal gauge invariants") under
the same roof. But this can only be done mitibation on the gauge side i.e. by
staying in a physical Hilbert space throughout the calculation.

Indeed some recent ideas about how to overcome this conceptual handicap go
precisely into this deirction. In order to have also physical (alias gauge
invariant) \textit{nonlocal operators within a unified formalism}, one must
leave the boundaries of gauge theory, because the latter by its very nature of
being a quantized form of classical gauge theories is limited to local
observables generated by pointlike fields.

There is a formulation for which free vectorpotentials are string-localized
$A_{\mu}(x,e)$ where $e$ is a spacelike direction. This potential is covariant
and fulfills the prerequisites of renormalization since its short distance
dimension is sdd=1. It is transversal and fulfills the axial gauge condition
$e^{\mu}A_{\mu}(x,e)=0$ in addition to transversality $\partial_{\mu}A^{\mu
}(x,e)$ \cite{MSY}$.$ One may call it the "axial gauge", but one should be
aware that strictly speaking \textit{it is not a gauge} but a covariant
string-localized field in the \textit{physical} Hilbert space which naturally
fluctuates in both variables $x$ and $e.$ With other words the direction $e$,
unlike a gauge parameter, participates also in the L-transformations and is
indistinguishable. from a point in 3-dim. de Sitter spacetime (space of
spacelike directions) and finally should also be accountable for string
localizations of charged fields (\ref{DJM})

Different from the pointlike setting of free potentials in the BRST (or any
other gauge fixing formalism), these covariant stringlike potentials share
together with their field strength the same physical Hilbert space and, at
least in the QED case, this continues in the presence of interactions. The
difference in the covariant transformation law requires to keep the
$e^{\prime}s$ (which participates in the Lorentz transformation) at generic
values, unlike a fixed gauge parameter in the pointlike BRST approach. Since
these stringlike potentials do not admit a Lagrangian description, one has to
take recourse to the setting of "causal perturbation theory". But this only
exists for poinrlike fields; the occurrance of stringlike localization leads
to a significant change in the perturbative iterative Epstein-Glaser formalism
which makes the perturbation theory different from those of pointlike gauge
fields in that counterterms may now also be string-localized. Ignoring this
aspect and treating it as a gauge problem in the axial gauge one inevitably
runs into the unmanageable infrared divergencies well-known to anybody who
tried to lay his hand on this problem. 

In the string-localized setting the origin of all these problems becomes
obvious since the infrared problems are equivalent to short distance problems
in a 3-dimensional de Sitter space, but unfortunately the problem does not
factorize in Minkowski and de Sitter, so that it necessitates a nontrivial
generalization of the Epstein-Glaser iteration step. This is presently beeing
investigated \cite{M-S}, but with only two people working on this problem,
(one being well beyond retirement age and the other overburdened with teaching
duties) this will take some time \cite{M-S}. One obvious observation should be
mentioned, for correlations of alias gauge invariant fields the new setting
leads to $e$-independence on the level of the same formal arguments as in the
BRST gauge formulation.

The renormalization theory involving string-localized fields is much more
demanding since the time-ordering does not only effect the starting points of
the semiinfinite strings, but also involves the string line as a whole (which
leads to the mentioned significant change in the Epstein Glaser iteration). It
is very important to do the computation for generic values of the $e_{i}$ i.e.
to treat them like independent points in de Sitter space and integrate, as one
always does, over the inner $x_{i}$.

The question is then whether one should average over the internal $e_{i}$
(integration) i.e. as if the theory would be a QFT on de Sitter space, or
whether one should smear all of them with the same testfuntion $g(e)$
supported around one point in de Sitter space which the above formula
(\ref{DJM}) would suggest. As long as one keeps the $g$ fixed on keeps the
terrible infrared problems of the axial gauge at bay. Most of our numerical
understanding about QCD comes from lattice analogs. But the use of lattice
theory is not such a good idea for problems of a more structural kind. Lattic
theory has not even been able to shed a light on the infraparticle problem,
how can it reasonably be expected to solve such structural conundrums as
invisibility in the sense of gluons, quarks or dark matter? 

The advantage of the formulation in terms of string-localized potentials
(instead of the standard formulation) is that the physical origin of the
infrared problems of QCD is clearer. But the problem is anything but simple,
and remembering how long ot took to get renormalization theory for pointlike
fields into a manageable shape, it would be unrealistic to expect that its
string analog can be worked out much faster than it took to elaborate
renormalized perturbation theory for pointlike fields.

One would like to expect from the string reformulation of nonabelian Yang
Mills theories some clarification of the following problems. The local degrees
of freedom, which can be described by pointlike physical fields, do not
account for all degree of freedom of the system. Wheras the fate of the
remaining one's in QED is well understood, in the QCD case this is terra
incognita and expected to account for the "invisible" degrees of freedom which
are carried by gluons and quark fields. Hence one would like to think that the
issue of invisible, inert or dark matter is connected to a very strong
indecomposable nonlocality beyond the well-known infraparticle properties
\cite{S3}\cite{S4} of undetected infrared photon clouds. The aim would be to
show that certain infrared degrees of freedom cannot be registerd at all in
counters which are at most quasilocal in their extension \cite{Haag}.

Even at standing accused of being repetitious let me state again that in
contrast to QM which can, by using appropriate potentials, keep matter "out of
sight" by placing it into a confining vault potential ("confinement"), the
only resource of QFT for creating its structural richness is causal locality;
there are no confining vaults in the arsenal. of causal localization. There is
of course no guaranty that properties as invisibility/darkness can be
explained within QFT, but there can be no doubt that the only available
resource is delocalization.

The analysis of irreducible representations has shown that there are two kinds
of localization, pointlike or indecomposable stringlike. In its purest and
strongest form the latter shows up in the stringlike fields associated with
the Wigner infinite spin representations \cite{MSY}. For arguments in favor of
their inertness relative to ordinary matter see \cite{S3}\cite{S4}. The
nonexistence of local operators which carry certain charge superselection
rules as in (\ref{DJM}) is only possible as a result of interactions.

A new string-localized formalism may also shed a new light on the
Schwinger\footnote{Schwinger was thinking of a screened "phase" in spinor QED,
where a perturbative implementation of screening is not possible. To make his
point more convincing, he invented the Schwinger model.}-Higgs formalism of
charge screening and its nonabelian counterpart. Algebraically there is no
difference between scalar QED (or its nonabelian counterpart) and the Higgs
model, since the presence of a degree 4 term in a would be charged complex
scalar field (needed for entering the "Mexican-hat" parameter region) is in
any case \textit{required by renormalization theor}y. In the presence of spin
1 fields, a charged Boson may \textit{screen itself} in the presence of
vectorpotentials and become a real field. The other degree of freedom in the
complex field together (in agreement with a structural screening theorem by
Swieca \cite{Bert}) with the two photon degrees of freedom combine to form a
massive vectormeson. 

At the end one has a fully pointlike local theory fitting into the standard
framework of QFT and instead of the complex massive field obeying a charge
selection rule the physical outcome consists in a real massive field without
any rule which limits its copious production. Does this have an intrinsic
meaning, can one experimentally tell that a model is the screened vversion of
an originally charged one ? Hard to say. In any case this fully local theory
is very different from the nonlocal electrically charged model, not to speak
about the even more nonlocal invisible hypothetical gluon degrees of freedom
in YM models.

The only conceptual recource which one has at one%
\'{}%
s disposal for the construction of interacting fields is Poincar\'{e}
covariance and locality. In perturbation theory one also needs a minimality
principle on the scaling degree, usually referred to as the renormalization
principle. Although constructions based on operator algebraic methods had some
recent success, the main source of qualitative and quantitative understanding
is still the local Poincar\'{e}-invariant coupling between free fields of
arbitrary spin and mass. Since all massive one-particle representations are
pointlike generated, there is good reason to believe (apart from the remark at
the end of the previous to last section) that, unless there are also zero mass
representations entering the coupling, the resulting theory will be generated
by pointlike fields. Adding global symmetries to these fields does not change
the localization properties 

Only the participation of higher helicity ($\geq1$) fields can do this. Power
counting reqirements in limiting the dimension of the interaction to
$sdd\leq4$ which excludes working with field strength and requires to use
instead their covariant potentials which turn out to be semiinfinite
string-localized. The problem is then to avoid that the whole theories is
becomes string-localized; each QFT should at least have a subset of (generally
composite) pointlike localized fields which generate the subalgebra of local
observables. This would be the analog of the gauge invariant algebra in the
approach built on local gauge invariance which the latter playing the role of
separating a physical content from an unphysical embedding. The local gauge
formulation not only misses out on "nonlocal gauge invariants" as the analogs
of the string-localized generators, but it also creates the wrong impression
that there is a mysterious gauge symmetry in analogy to an inner compact group
symmetry, whereas it is really the existence requirement of a nontrivial local
observables which restricts the interactions beween stringlike potentials and
their coupling to massive matter. Nothing is known about such nonlocal degrees
of freedom and their expected lsck of "visibility" beyond that related to the
photon inclusivness in QED. It would be intereting if the present unparticle
activities could be directed towards these gaping holes in our understanding.

An indication of non-intrinsicness comes from the computation where one starts
from the sdd=2 massive vectormeson, uses BRST to lower the dimension to ssd=1
(which leads to renormalizability by power counting) and requires that the
BRST cohomological formalism also works in higher orders. It turns out that
this can only be achieved by introducing an additional physical degree of
freedom whose simplest realization is a scalar massive particle (naturally
with a vanishing one-point function which is the hallmark of the Higgs). But
there is an unsatisfactory aspect in such a derivation since cohomological
requirements in a an indefinite metric space are not representing physical principles.

The use of "ghostly crutches" could be avoided by using a string description
of a massive vectormeson field which does the same job as BRST (namely
reducing the sdd to 1 \cite{MSY}) without introducing indefinite metric and
remaining in the physical Wigner-Fock space. In that case the only remaining
principle for the presence of an additional scalar particle is that without
its presence there would be a problem with locality i.e. the use of a
string-localized vectormeson has made the interaction renormalizable in the
sense of power counting and in order not to remain stuck with only
semiinfinite strings we would need a locality restoring scalar particle. If
the strings can be resoled in terms of pointlike fields without the presence
of an additional scalar then we would have learned that there are fully local
vectormeson theories without scalar companions. No matter what LHC will tell
us, both outcomes would have fundamental consequences for the development of
QFT. Arguments in favor of higher spin particles which only can interact among
each other in the presence of \ lower spin particles would be much deeper than
those based on higher symmetries as e.g. supersymmetry. Only in this way can
one get away from the Mexican hat cooking recipe for the Schwinger-Higgs
screening mechanism

\section{What are unparticles and are they related to infraparticles ?}

According to the existing literature unparticles \cite{Geo}\cite{Geo2}%
\cite{Grin} are hypothetical bursts of scale invariant invisible "stuff" which
is formed in high energy collisions of ordinary particles and which dissipate
without leaving direct traces ("invisibility") through secondary interactions.
Whereas the unobserved infrared photon "stuff" below the resolution leads to
inclusive cross sections and changes the nature of the charged particles in a
conceptually very radical way, the idea around unparticles is different namely
after pealing off some unobserved long range "stuff" into the vacuum, the
source particles remain the same ordinary particles as before the interaction.
Such a process appears somewhat strange. since a short range source should not
be able to give rise to "pealing off" long range stuff unless this stuff
consists of massless particles but not as members of an infinite cloud. An
example of such an reaction would be the before mentioned infrared-tame
interaction of massive spinors with scalar massless (and quite visible) mesons.

In order to avoid getting lost in vagueness, we first look at the concrete and
explicit 2-dim. illustration in \cite{Geo2} which consists of the modified
Schwinger model i.e. $QED_{2}~$with a massive vectormeson instead of photon.
As mentioned in the previous section all soluble models of the 60s
\cite{Elcio}, apart from the original Schwinger model itself, contain the
subtle charge-carrying exponential Boson factor and other standard free
fields; this is also the case for the modified Schwinger model. The authors
chose for their illustration the chiral condensate operator \cite{L-S} which
for the modified model has the form (here the details of how this operator
comes about from a Lagrangian are irrelevant)%
\begin{equation}
\mathcal{O}(x)=:e^{i\alpha A(x)}::e^{i\beta\varphi(x)}:
\end{equation}
where $A(x)$ is the gauge invariant massive field the exponential of which
already appears in the gauge invariant solution of the Schwinger model. The
second exponential is a charge carrying zero mass terms and $\alpha$ and
$\beta$ are real parameter related to the ratio of the mass coming from the
Schwinger-Higgs mechanism and the Lagrangian vectormeson mass $m_{0}.$ If it
would not be for the second factor, the large distance limit would describe
the Schwinger-Higgs chiral condensate with the massive one particle
contribution being the next leading term in the expansion. The presence of the
charge-carrying massless exponential undoes part of the screening and converts
the leading term into a "infravacuum" wheras the next to leading term
represents an infraparticle contribution in the previously explained
sense\footnote{A footnote in \cite{Geo2} reveals that the authors are aware of
this connection.}. This charged "infravacuum"\footnote{The quotation marks are
there in order to distinguish this situation from a more radical notion of
infravacuum \cite{Ku} which cannot be viewed as the application of a charge
carrying zero mass field to the standard vacuum. .} component of
$\mathcal{O}\Omega~$is the only component which resembles separate
scale-invariant "stuff" similar to what the authors envisage for unparticles;
but try to have an interacting situation in which conformally invariant
components coexist with massive ones in d=1+3 and watch yourself failing; to
talk about a sector which is a little bit nonconformal is not much better than
introducing the notion of a little bit pregnant in real life.

Two-dimensional models of the mentioned kind do not describe scattering.. Even
though they are not free fields in the technical sense, they describe
noninteracting charged "stuff". So in order to utilize the infrared
contributions to the two-point function in Feynman diagrams, the authors
couple $\mathcal{O}$ to the square of another field \cite{L-S}. They use the
fact that the infra/unparticle structure in d=1+1 allows for interaction-free
illustrations (which only look like containing interactions) whereas according
to our discussion in the previous section it is not possible in d=1+3 to
separate kinematical from dynamical aspects.

Whereas d=1+1 infraparticles were introduced in order to understand the
scattering of charge particles, the unparticles in the sense of representing
scale invariant "stuff" which, unlike the soft photons clouds which never
liberate themselves from the charged particles, are apparently not hooked on
massive matter. Accepting for the sake of the argument the properties their
protagonsists like to attribute to them it seems that they do not appear in
the outgoing amplitudes and are not even accounted for in inclusive cross
sections. It seems that the example of the extended Schwinger model (as all
other examples with coupling to massless scalars) is not a good illustration
for the creation of zero mass conformal stuff which, unlike that of
infraparticles and its zero mass clouds, is supposed to separate itself from
ordinary massive matter.

This model also points at two unsolved problems in QFT. The first one is: does
it make sense to couple free fields with fields which are interacting from the
start? Besides the question of practicality for a perturbative approach there
looms an unsolved fundamental problem. One formulates interactions by coupling
free fields not only for pragmatic computational reasons. One also believes
that this insures the mentioned asymptotic completeness, which in the mass
zero case amounts to a weaker form of completeness in terms of inclusive cross
section. But it is doubtful that the coupling of anomalous dimensional
conformal matter coupled to free fields stays in this setting.

The second difficulty which becomes particularly acute in d=1+3 is that
coupling of massive to massless matter never leads to scale-invariant
"sectors"; the only theory which does this is the tensor product of a
conformal theory with a massive one. As much as it is meaningless to use
expressions as "a little bit pregnant" in daily life, one cannot fight
structural properties of QFT by notions of effective field theories or what is
more specific to the situation at hand by Bank-Saks arguments which claim that
it is possible to overrule such structural facts and make sense out of
violating conformal invariance in a region of a theory. Ideas of effective
actions may have their place of validity, but one should not try to use them
for overturning structural properties.

As the example of the photon shows, its interpolating Heisenberg field is not
scale covariant, only the registered outgoing free photons are. What remains
however intact is the gapless zero mass energy-momentum spectrum and its
ensuing long range character. It is also interesting to point out that even on
a formal level the coupling of anomalous dimensional fields with ordinary
matter does not improve the long range aspect; to the contrary, as the scale
dimension increased, the infrared coupling becomes weaker. The strongest
infrared couplings are those which involve string-localized potential
associated with ($m=0,s\geq1$) representations for which the aforementioned
vectorpotential $A_{\mu}(x,e)$ is the best studied case. The use of the
string-localized description makes the long range which sets the infrared
strength of the QED coupling manifest whereas (see previous section) in the
gauge formulation this remains hidden and has to be brought out "by hand"
through formulas as (\ref{DJM}). Only they have a chance for accounting for
the desired invisibility property.

Without wanting to lend support to the somewhat controversial physical
interpretation of such couplings in the literature on unparticles, it may be
interesting to mention that there are anomalous dimensional conformal fields
which describe "stuff" in a more literal sense i.e. something which
\textit{certainly cannot be interpreted} as coming from a scaling limit from a
standard theory and therefore cannot be associated with inclusive cross
sections. These are the \textit{conformal generalized free fields} as they
e.g. arise from ordinary AdS free fields via the AdS-CFT correspondence. From
a combinatorial point of view they behave as free fields\footnote{They are not
the standard anomalous dimensional fields which are "interacting" in some
sense which can be made precise.}; Duetsch and Rehren \cite{Du-Re} have
investigated the suitability of the causality properties of such "stuff" for
the formulation of a consistent perturbation theory and their results. The
results are yet incomplete, but interesting and even somewhat encouraging.
None of the unparticle lowest order calculations which only use unparticle
two-point functions would change, if one uses these combinatorially much
simpler fields.

Perhaps one should be careful with prematurely attaching physical attributes
to unparticle calculations and rather study in more general terms what QFT has
still in store once one goes beyond standard textbook particle physics. The
most fruitful unexplored area seems to be the afore mentioned interacting
massless higher helicity objects coupled to themselves and to standard massive
matter. It is the generalization of the mentioned string-localized
electromagnetic vector potential with scale dimension sdd=1 which has a good
chance to lead to the kind of infrared singular interactions which one needs
to get beyond the standard matter and create "stuff" which consists of
infinitely many objects in a finite energy range. Zero mass is necessary but
not sufficient; e.g. scalar zero mass couplings do not have the sufficient
infrared strength. It is not the size of anomalous dimension but rather the
algebraic form of the infrared coupling of higher helicity free
string-localized potentials which increases with spin that increases this strength.

Among all at least partially studied models, the most promising are those
which involve couplings among several string-localized potentials $A_{\mu
}^{(i)}(x,e).$ The experience with nonabelian gauge theories suggest that in
order to find pointlike generated subalgebras the couplings must be related to
each other in the way they are in gauge theories; if not one will get stuck
with a string-localized theory which has no local subobservables at all.
Assuming that the mentioned string theoretic generalization of renormalized
perturbation theory works, one would have two kind of matter in such a
setting: visible point-localized "glue-ball" matter and string-localized and
presumably invisible gluon matter. I am convinced that without solving this
problem one has no chance to understand the issue of invisibility versus
asymptotic incompleteness. The understanding of the abelian counterpart QED
where such string-localized fields represent the charged operators and where a
rest of undetected soft photon "stuff" always remains outside observation is
encouraging for a program of looking for stronger forms of invisibility. The
unparticle project certainly shares this aim even if the proposals to
implement it are quite different (apart from the shared low-dimensional illustrations).

The unparticle project tries to achieve invisibility of interaction generated
"stuff" by using conformal matter with anomalous dimensions, in contrast the
project favoured in this article is based on a generalization of gauge theory.
Whereas the gauge formulation hides the nonlocality by introducing fake
pointlike potentials together with BRST ghosts at the expense of the Hilbert
space positivity and as a result tends to overlook (even in the abelian case
(\ref{DJM})) nonlocal operators in the \textit{physical} Hilbert space, the
string like description catches also those nonlocal field degrees of freedom
which escape the pointlike description but nevertheless carry energy-momentum
and hance react gravitationally. This still speculative project which
generalizes the infraparticle idea is expected to explain the confinement of
the gluon and quark degrees of freedom and to attribute physical reality to
genuinly invisible dark matter. In the concluding section I will address this
speculative issue in a more general context than un/infra particles.

\section{Invisibility and lack of asymptotic completeness}

Whereas in QM, which has no maximal velocity, fields are synonymous with
particles and there is hardly any limit on the kind of interactions between
them, QFT is more restrictive as a result that all of its properties at the
end must be understood in terms of causal localization i.e. the localization
in theories which have a maxial velocity \cite{interface}. The prize to pay
for this is that its only measurable non-fleeting and genuinly intrinsic and
stable objects, the particle states, are only appearing in the large time
asymptotic limit of the fleeting field states respectively; an interacting
theory with any interaction fulfilling the general principles will have no
particles at any finite times! Since besides the stability (the existence of a
lowest energy state) the realization of causal locality is the only handle at
one's disposal in order to control the asymptotic particle content, the study
of admissible particle structure and their possible manifestations in the real
world has remained the most subtle part of QFT.

Even the deeper understanding of the standard situation, in which one-particle
states are separated from the continuum by a gap, has remained a 50 year
challenge \cite{Bert}. It started with the (at that time surprising)
observation that the number of phase space degrees of freedom in a finite
phase space cell (which as everybody learns is finite in QM), is infinite in
QFT; an infinity which originates from the realization of the causal
localization principle. The hope was that the precise quantitative
understanding of this infinity could explain why the Hilbert space of QFT
apparently can be fully described (even beyond perturbation theory) as a
Wigner-Fock Hilbert space; a fact which ceises to be valid in QED.

For free fields this set is compact and (as was shown later) even nuclear
\cite{Haag} and there are good arguments that at least in physically
reasonable theories (e.g. absence of Hagedorn temperature) it stays this way.
Although many deep properties followed from this phase space structure, it is
now agreed on that asymptotic particle completeness cannot be derived from
phase space properties alone. 

The infraparticle structure of electrically charged particles contained the
important message that there are objects which cannot be generated by
pointlike field and the formalism of gauge theories. Nonlocal objects in the
physical Hilbert space as e.g. electrically charged particles are better
constructed in a setting which permits stringlike localized potentials
\footnote{From a point of view of positive energy irreducible representations
of the Poincare group, the necessity to introduce stringlike generators only
arises for the zero mass potentials of the pointlike helicity $\geq1$ field
strength and in a much stronger form for the so-called infinite spin
representations (which possess no pointlike generators). There is no reason
which forces one to go generators on higher dimensional submanifolds as
branes.}.

It is interesting to note that as long as the mass gap hypothesis holds, the
formulation and derivation of scattering theory between pointlike and
semiinfinite stringlike fields \cite{Haag} is similar. The only significant
difference is that the S-matrix and the formfactors of such models do not
necessarily fulfill the important crossing property \cite{foun}; with other
words there is no analytic master funcrion such that the different
distributions of in and out particles are different boundary values of that
master function. This has the intersting consequence that there may be a
subterfuge to Aks theorem \cite{S3}\cite{S4}. In that case it would be
possible that certain channels cannot be produced in scattering processes
despite the fact that there is no charge superselection rule which prevents
them. this kind of partial inertness may have some potential interest in
connection with dark matter. Massive strings can only exist in interacting
matter, free massive fields are always pointlike generated. Massive strings
once applied to the vacuum create states which are always generated in terms
of pointlike states (which themselves cannot be obtained by applying pointlike
fields from the interacting field algebra).

From the existing formulation of the unparticle project it is not clear if and
how they fit into the balance of asymptotic completeness. For standard mass
gap situation of believes that the coupling between free fields is not only
popular because it is simple and one cannot think of anything else which is in
agreement with the locality principle, but it also preempts the property of
asymptotic completeness by having from the very beginning those fields and
their Fock space which are as close as possible to the incoming/outgoing
fields. In the case of electrically charged fields though for infraparticles
one looses the Fock space structure, but since the scattering probabilities
still add up (the resolution can be made arbitrarily small). In the case of
QCD and for unparticles this is not so clear, but for different reasons. The
difficulty in the latter case is that one starts already with anomalous
dimensional fields which are not only very far removed from a Fock space
structure, but for which it is not even clear whether they permit a doubly
inclusive cross section setting which appears to be the least one needs to
fulfill asymptotic completeness in the sense of probabilities.

\section{Concluding remarks, resum\'{e}}

The unparticle idea presents an opportunity to recall previous successful and
less successful works which explored the region beyond the standard (mass gap)
particle setting. The theory of infraparticles which aimed at the
incorporation of the observed infrared aspects of electrically charged
particles is an example of an successful attempt. Apart from the issue of
"invisibility" which is the main motivation behind unparticles as a new kind
of matter, the infraparticle models follow a similar construction recipe as
those designed to illustrate unparticles, in fact \textit{its two-dimensional
illustrations are identical}.

On the other hand the popularity of unparticles may point our thinking again
towards invisibility problems caused by noncompactly localized degrees of
freedom within theories which we prematurely believed to have "solved" (e.g.
by analogies with lattice theories). Some nice catch words as (gluon-, quark-)
confinement provided us with a quiet conscience. But lattice theory lacks
those strong principles which relate indecomposable positive energy
representation of the Poincar\'{e} group for certain zero mass representation
to semiinfinite string localization\footnote{The problems of lattice
approximatio of QFT is similar to the approximatabilty of operator algebras by
matrix algebras. Although for hyperfinite von Neumann algebras this is
possible, there is no way of keeping track of the richness of infinite class
of hyperfinite algebras by looking at finite matrix algebras. Most of the
interesing physical mechanism are only accounted for in the infinite limit.},
although it is quite efficient at emulating standard QFT containing only
compactly localized representations. Interaction-free illustrations exist in
the form of string-localized infinite spin positive energy representations in
Wigner%
\'{}%
s list, but unfortunately they do not generate local subalgebras (they have no
local conserved energy-stress tensor), at least not in this interaction-free form.

What one needs in order to have states "out of sight" coexisting with states
generated by local observables is a situation in which the interaction is so
strong in the infrared, that besides ordinary matter (described by a local
subalgebra) there are degrees of freedom (not ghosts!) which have a
localization as bad as that of the mentioned infinite spin representations.

The existence of ordinary matter in QCD-like theories has become part of the
accepted folklore (dimensional transmutation) and phenomenological schemes
have been designed to link such mechanisms with Lagrangian QCD. But the other
side of the coin, namely the fate of the nonlocal (gluonic) degrees of freedom
which escape the gauge theoretic formalism (which by its very nature is only
focussed to the local ones) have been left in a conceptual limbo.

Looking at the literature, it seems that most people believe that they do not
exist, i.e. that all the degrees of freedom went via dimensional transmutation
into the local observables which in turn form an asymptotically complete
system. But in view of the electrically charged particles whose sharpest
possible localization is semiinfinite string-like, this is not very credible.

In order to see what is going on, we have started to investigate quadrilinear
interactions between stringlike free fields with special attention to those
for which the string-like nature is not enforced by imposing renormalizability
but is naturally emerging from the Wigner representation struture
\cite{MSY}\cite{M}\cite{M-S}. Since such fields have sdd=1 independent of
spin, there is no problem with the power counting prerequiste for
renormalizability. The really hard problem, as mentioned before, is the
perturbative Epstein-Glaser iteration for semiinfinite strings instead of
pointlike particles \cite{M-S}.

There is no disagreement with the aims of unparticle physics since almost all
the deep unsolved problems are in the infrared. Whatever the outcome will be,
it is important to get particle physics away from those metaphoric inventions
as e.g. supersymmetry and string theory back on track addressing the old
unsolved problems with new ideas.

My attitude with respect to unparticles has been one of criticism (mainly
connceted to the knowledge which was lost during the last 3 decades) but at
the same time encouragement because unparticles could serve as a catalyzer for
a return to particle theory's most rich research area, which, unfortunately
left too many unsolved problems on the wayside\footnote{Inasmuch I have
lamented the loss of criticism in another context, I am of course also aware
of the danger of calling an idea to account for conceptual-mathematical rigor
in a too early stage; there are several important ideas in particle physics
which started out on a wrong track. }. One central problem which was first
formulated in the early 60s \cite{Swieca} is: \textit{when is a theory of
quantized fields a theory of particles}, or more specifically. the problem of
asymptotic completeness. The phase space structure of QFT (which is
significantly different from that of QM) was identified as one local
structural property which plays an important role in the understanding of the
global particle structure. It was already clear at that time that there are
field models which do not fulfill asymptotic completeness (generalized free
fields, conformal models) and they should be excluded. But physically
important structures as electrically charged particles and more general
infraparticles are still inside an appropriately extended asymptotic
completeness notion. Their history shows that structural problems of QFT
cannot be solved by lowest order perturbation theory; the first order coupling
between a conformal field theory and massive matter evaluated for the two
point function, as used by unparticle followers, does not reveal anything. 

It is far from being clear why, what was considered to be an important
physical principle or at least a successful working hypothesis in those times,
should be ignored now. After all there is presently not the slightest
indication that nature does not like the old principles nor is there a
theoretical guide outside of at least some form of asymptotic completeness in
the sense of probabilty conservation when unparticle "stuff" fades away into
the vacuum.

\end{document}